# CSCR: Computer Supported Collaborative Research

Vita Hinze-Hoare
School of Electronics and Computer Science
University of Southampton
United Kingdom
2006

**Abstract**
*It is suggested that a new area of CSCR (Computer Supported Collaborative Research) is distinguished from CSCW and CSCL and that the demarcation between the three areas could do with greater clarification and prescription.*

## 1. Introduction

The twin fields of Computer Supported Collaborative Work (CSCW) and Computer supported Collaborative Learning (CSCL) have been the subject of intense interest in the HCI research community during the past seven years.

The split between CSCW and CSCL has grown wider in response to the recognition that the learning process is more distinct from the working pattern and is more intensively understood through new theories of pedagogy and education.

It has become apparent that CSCL requires all of the facets of CSCW but in addition is constraint by these pedagogical theories and as such it is argued here that CSCL is a subset of CSCW (see figure1)

The process of research is also a learning process but one which is more highly refined and involves learning in a particular way using special techniques and tools. As such it is argued further that research which is supported by computer collaboration is a subset of CSCL (fig.1)

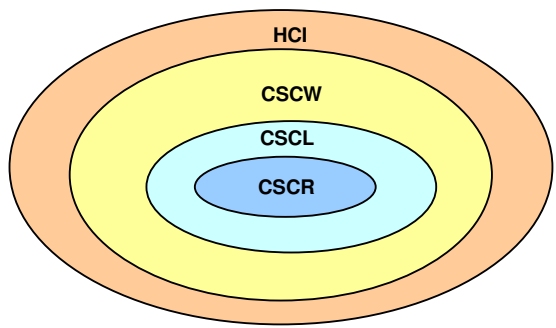

**Figure 1**

## 2. Differences between CSCW and CSCL

Diaper (2005) maintains that the History of HCI shows a lack of coherent development. There is no agreement as to what HCI is, should be, or do. The discipline is becoming increasingly fragmented to the point where it is difficult to establish consensus in the field. This fragmentation of discipline of HCI is already so extensive according to Diaper that it is hard to even characterise the method of approach.

Much the same is true of CSCW and CSCL. These have been the subject of extensive research for a number of years there is still no accepted definition of either. "This lack of agreement highlights the necessity for the development of a general systems model, both in the general HCI approach and in the specific collaborative approach" Diaper (2005)

The primary difference between CSCW and CSCL is that CSCW can be characterised by the need for "workingspace", while CSCL needs both "workingspace" and "Learningspace".

WorkingSpace is that domain where the following activities can take place

- **Communication Space**
  A set of interactive tools common to CSCW would be required to maintain communication and the interchange of ideas in real time.
- **Scheduling Space**
  Collaboration requires both synchronous and asynchronous communication and scheduling of meetings, the setting of deadlines, setting up of conferences (online or otherwise). A common scheduling facility is required to maintain collaborative structure.
- **Sharing Space**
  Collaborative research necessitates the exchange of information which may be in multimedia formats such as sound, video, image text etc.
- **Product space**
  Knowledge artefacts are the expected outcome of the learning process and a tally of these need to be kept and maintained as a record of work done and an indication of progress

LearningSpace is that domain in which all the aspects of workingspace are available and in addition the following can take place.

- **Reflection space**
  An important part of learning which has been recognised by recent pedagogists is the need for internal reflection. This can be both individual and collaborative and could be assisted with the help of an on-line journal (Private and Group)
- **Social space**
  Much learning has been shown to arise from interaction with peers and other learners as well as from a didactic intercourse with mentors. Daniels H. (2001)
- **Assessment space**
  The learning process needs ratification through a testing regime. This may involve the provision of online questions and assessment.
- **Tutor space**
  The dual roles of teacher and learner need to be reflected in the construction of a CSCL environment. Tutors would require their own private area for their specific tasks.
- **Administration space**
  The day to day management of course data and the administration of learning tasks, student information will require its own area.

| The Spaces required by each of the collaborative areas |||
|---|---|---|
| **CSCW WorkingSpace** | **CSCL LearningSpace** | **CSCR ResearchSpace** |
| Communication | - | - |
| Scheduling | - | - |
| Sharing | - | - |
| Product | - | - |
|  | Reflection | - |
|  | Social | - |
|  | Assessment | - |
|  | Tutor | - |
|  | Administration | - |
|  |  | Knowledge |
|  |  | Privacy |
|  |  | Public |
|  |  | Negotiation |
|  |  | Publication |

**Figure 2**

## 3. Differences between CSCL and CSCR

The primary difference between CSCR and CSCL is that a complete record of all interactions between participants is an important and necessary tool to evaluate the contributions of each member in a collaboration group which can later on determine "a fair capital share" if the undergoing research project is successful. Further differences will include the following

- **Knowledge Space**
  Research collaboration will generate its own knowledge base and a database system will be required which can store and retrieve this information as well as allocating ownership to individual contributions to ensure an appropriate apportionment of credit. It would be expected that this system would incorporate hypertext and links to bring cohesion to individual contributions.
- **Publication space**
  The publication of pre-prints and draft papers to online sites such as arxiv.org would be assisted by an automated process incorporated into the system.
- **Privacy Space**
  The research group will need a private area in which to work that is closed to non-group members. It is important to maintain a secure area where work is developed before it is published.
- **Public Space**
  The collaborative research group may wish to provide information upon the nature of the research which is being done, to encourage contributions, questions, raise issues etc. which can be place online in the public domain.
- **Negotiation Space**
  Group research may often introduce conflicts of opinion which need to be worked through. This is more difficult online and may involve intensive and protracted discussions.

## 4. Conclusion

Unlike CSCW and CSCL it has been shown that there is a case to be made for regarding CSCR as a separate and distinct area of investigation. Each of these domains has their own specification and requirements. The first two have much more than that including their own "conferences, journals and adherence" Stahl, G. (2003). The latter is yet to develop and is a potential fruitful area for future research. All three domains have commonality and dependency, and borrow from one another. However, CSCR has individual aspects which are not part of the other two, and consequently is distinct and should be treated as such.